\documentclass{osa-article}
\usepackage{subcaption}
\usepackage{amsmath}

\usepackage{graphicx}
\usepackage{subfiles}
\usepackage[ruled,vlined]{algorithm2e}
\usepackage{multirow}
\journal{osac}

\articletype{Research Article}
\usepackage{cite}
\usepackage{lineno}
\linenumbers

\begin{document}
\nolinenumbers
\title{Weight Bank Addition Photonic Accelerator for Artificial Intelligence }

\author{Wenwen Zhang,\authormark{1}Hao Zhang\authormark{1,2}}

\address{\authormark{1}Department of Electric Engineering and Computer Science, University of British Columbia, BC, Canada}
\address{\authormark{2}Department of Electric Engineering and Computer Science, University of Victoria, BC, Canada}



\begin{abstract}
Neural networks powered by artificial intelligence (AI) play a pivotal role in current estimation and classification applications due to the escalating computational demands of evolving deep learning systems. The hindrances posed by existing computational limitations threaten to impede the further progression of these neural networks. In response to these issues, we propose neuromorphic networks founded on photonics that offer superior processing speed than electronic counterparts, thereby enhancing support for real-time, three-dimensional (3D), and virtual reality (VR) applications. The 'weight bank'—an integral component of these networks—has a direct bearing on their overall performance. Our study demonstrates the implementation of a weight bank utilizing parallelly cascaded micro-ring resonators (MRRs). We present our observations on neuromorphic networks based on silicon on insulators (SOI), where cascaded MRRs play a crucial role in mitigating inter-channel and intra-channel cross-talk, a persistent issue in wavelength division multiplexing (WDM) systems. Additionally, we design a standard silicon photonic accelerator to perform weight addition. Optimized to offer increased speed and reduced energy consumption, this photonic accelerator ensures comparable processing power to electronic devices.

\end{abstract}
\section{Introduction}
Communication demands proliferate in recent years with growing data volume and expanding data capacity. The popular industry needs higher processing speed and more extended bandwidth to make the connection faster and more stable. And in the emergence of the internet of everything (IoT) and intelligent embedded society, tremendous data analytical tasks are requiring highly-computational ability with better energy efficiency. Traditional electronic technology is reaching the limit in providing larger processing bandwidth and lower energy consumption, where integrated photonics devices play an important role\cite{Tait:18}. In the post-Moore era, the photonic accelerator becomes feasible by making use of their frontier novel nanostructured to reach goals of miniaturization and power efficient \cite{ kitayama2019novel,shen2017deep }. Silicon photonic devices perform better than other photonic devices in terms of integration, compactivity, response to tuning through thermo-optic effect \cite{perez2017multipurpose,komma2012thermo} and reconfigurable linear systems \cite{liu2016fully,harris2016large,perez2017multipurpose}. Recently, a rapidly growing interest in neuromorphic photonics has been arousing for the potential possibility for machine intelligence by cooperating ultra-fast speed of photonics with the high energy efficiency of neuromorphic structure \cite{prucnal2017neuromorphic,tait2017neuromorphic,de2017progress,shastri2017principles}. A recurrent network is significantly useful in many applications at a smaller scale than fully connected networks or deep networks since some neurons can be reused \cite{ 8919993}. \cite{ ma2020blind} put up problems with management over ubiquitous interferences between channels and widespread heterogeneous wireless broadcasting. They propose a method for de-mixing mixed signals and decoupling photonic blind source separation (BSS) with frequency-dependent knowledge of the target frequency band. Other research also shows pave the way for advanced blind source separation in multi-antenna/array systems can greatly exceed current approaches based on digitizing ultra-redundant multi-dimensional signals \cite{russer2004signal}.

Micro-ring resonator (MRR) weight banks make silicon photonics weighted addition possible. Tunable MRR weight bank based on a wavelength division multiplexed (WDM) optical signal has been proven to vastly exceed the capabilities of electronics process analogy signal in photonics \cite{Tait:18,tait2017neuromorphic}. Photonic principal component analysis (PCA) approach which makes high-performance dimensionality reduction in wideband analog systems possible is also realized by weight banks, configuring record-high accuracy and precision weight banks and generating multi-channel correlated input signals in a controllable manner \cite{ma2019photonic,tait2019demonstration,tait2015demonstration}. Independent component analysis (ICA) is also applied to photonic structures by on-chip MRRs to reveal underlying independent but hidden factors in mixed signals. \cite{ma2020photonic}. However, inter-channel and intra-channel crosstalk is a significant source of signal degradation in WDM systems which reduces WDM channel count, increases insertion loss, and degrades adjacent channel isolation. Cascaded and series-coupled (second-order) MRR \cite{tait2016continuous,tait2016multi,tait2016silicon} filters are proven to be effective for low intra-channel crosstalk at higher data rates by offering larger input-to-through suppression over wide bandwidths \cite{jayatilleka2016crosstalk}. To expand the usable channel number and extend the free spectral range (FSR) of MRR filters, a grating-assisted coupler has been applied to MRR filters as well \cite{mistry2018bandwidth}. In this work, we applied series a of standard photonic accelerators based on coupled MRRs weight banks.

\begin{figure}
\centering
\includegraphics[width=0.6\textwidth]{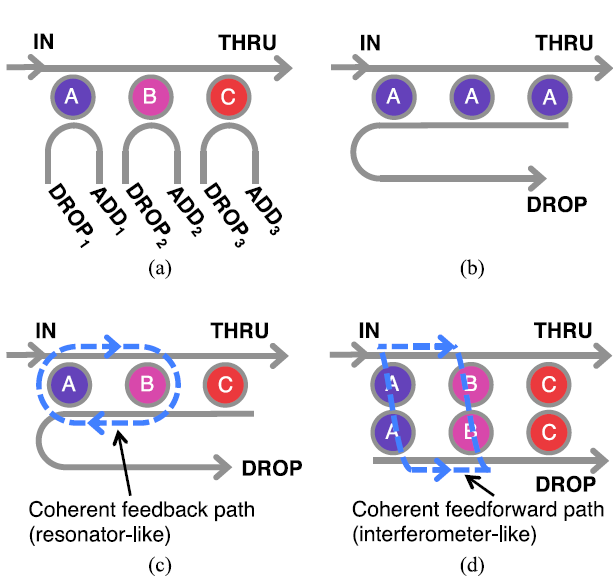}
\caption{\label{fig:weight bank 4} Different types of weight bank \cite{tait2016microring}. (a). Add-drop multiplexer. (b). Dual-band double channel side-coupled integrated spaced sequence of resonators (SCISSORs). (c). 1-pole MRR filters. Each MRR controls a separate WDM channel. Two waveguides make coherent feedback between surrounding MRRs. (d). 2-pole MRR filters. Interferometer-like feedforward coherent interactions. A B and C letters represent different WDM channels affected by the appointed resonator. }
\end{figure}

Several ways to interconnect photonic neurons have been developed based on MRRs. MRRs are the most intuitive structures to conduct tunable weights in neuron networks because MRRs can direct signals to through/drop ports by simply changing resonance frequency. This makes plus/minus weighing easy to control. Inter-channel cross-talk will largely reduce the channel space and channel counts. The research focused on inter-channel cross talks try variant ways to suppress inter-channel cross-talk and increase circuit efficiency. 

MRRs need waveguides to a couple in and a couple out optical lights, and there exist different ways of coupling \cite{ tait2016microring }. In Fig. \ref{fig:weight bank 4}(a), each MRR is coupled to a distinct and parallel waveguide in the drop port, creating coherent interactions between neighboring MRRs. Cross talk occurs if the wrong channel is coupled partially and runs into an incorrect drop port. Dual-channel side-coupled integrated spaced sequences of optical resonators (SCISSOR) are introduced and analyzed in \cite{mancinelli2011optical}, as shown in Fig. \ref{fig:weight bank 4}(b). Let each MRR in Fig. \ref{fig:weight bank 4}(b) control the weight of the separate WDM channel. Every channel is coupled to a neighboring MRR, which can decrease inter-channel cross-talk and create a feedback path for multiple MRRs, as is shown in Fig. \ref{fig:weight bank 4}(c). When using series or cascade (even poles) MRRs as components, corresponding weight banks create feedforward paths with coherence. Interactions between multiple MRRs should be considered for channel density problem, as is displayed in Fig. \ref{fig:weight bank 4}(d).

\section{Methods} 
\subsection{Device fabrication and characteristics}
This silicon photonics chip is fabricated based AMF technology using silicon-on-insulator (SOI) wafer at Institute of Microelectronics (IME) A*STAR foundry \cite{lim2013review}. The silicon wafer platform is $200mm$. The 220nm-height silicon is buried in oxide with $2\mu m$ thickness. To fabricate low-loss waveguide with $500nm$ width, $248nm$ \& $193nm$ deep ultraviolet (DUV) lithography is adapted. One 4*4 MRR weight bank is composed of 4 MRRs in parallel/series cascaded by 2 bus waveguides (<1.5 dB/cm) to trap light around these MRRs. TE mode lights directed by 8 degree grating couplers are tested in this fabrication with around 4.8dB insertion loss. The output is detected by balanced photo-detectors (BPDs). Mental routing traces and vias are also deposited for interconnects and for electrical probe/bond pads. Individual mental pads for each MRR are probed to thermally tune the resonance performance while the mental pad for ground effect (GND) are shared by all MRRs to make the whole chips neater and reduce electrical I/O ports.  

\subsection{experimental setup and testing}

\begin{figure}
\centering
\includegraphics[width=0.8\textwidth]{ 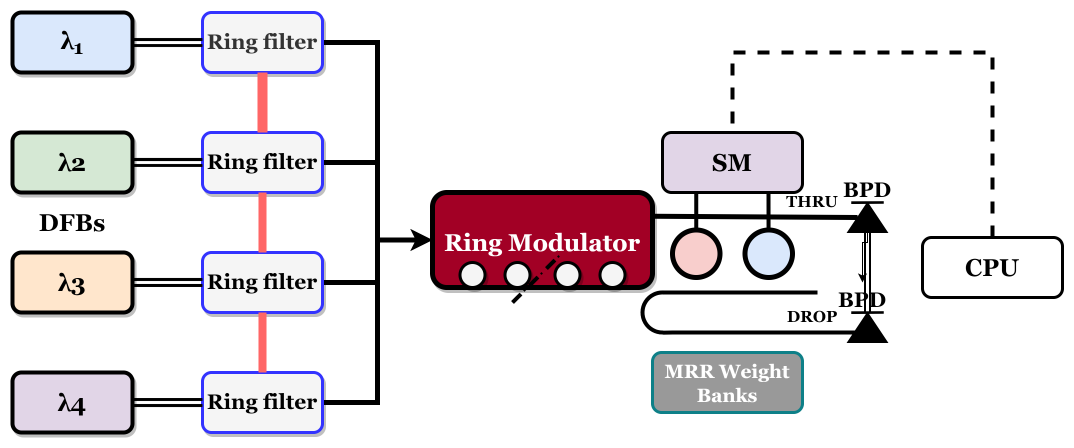}
\caption{\label{fig:diagram} Schematic of the experimental setup for performing photonic accelerator using an on-chip MRR weight bank. DFBs: distributed feedback lasers. SM: source meter. MRR: micro-ring resonator. BPD: balanced photo-detector.}	
\end{figure}
The expected experimental setup is shown in Fig. \ref{fig:diagram}. Distributed feedback lasers (DFBs) generating optical carriers at 1550 nm provide optical input signal for the whole photonic accelerator. The optical carrier then flows into four ring filters and then four ring modulators. Input optical carriers are then multiplexed together after these ring resonators. These optical carriers then experience wavelength-dependent channel delays under the effect of two split y-branches with different optical path lengths to construct partially correlated inputs. Eventually optical inputs enter the IN port of the MRR weight bank where weighting on the silicon photonic chip happens.

For tunable purpose, MRR weight bank is mounted on temperature-controlled fiber alignment stage. The insertion loss at measurement will include grating coupler insertion loss (about 5 dB), and waveguide/bending loss (around 1 dB). The OUT signal at THRU and DROP ports of the MRR weight addition are summed up by on-chip balanced photo-detector (BPD). For thermally adjusting, MRR will be cooperated with N-doped ring-shape heater which is driven by a source meter set in current-source, voltage-measure mode.

\section{Weight bank design}
\begin{figure}
\centering
\includegraphics[width=1\textwidth]{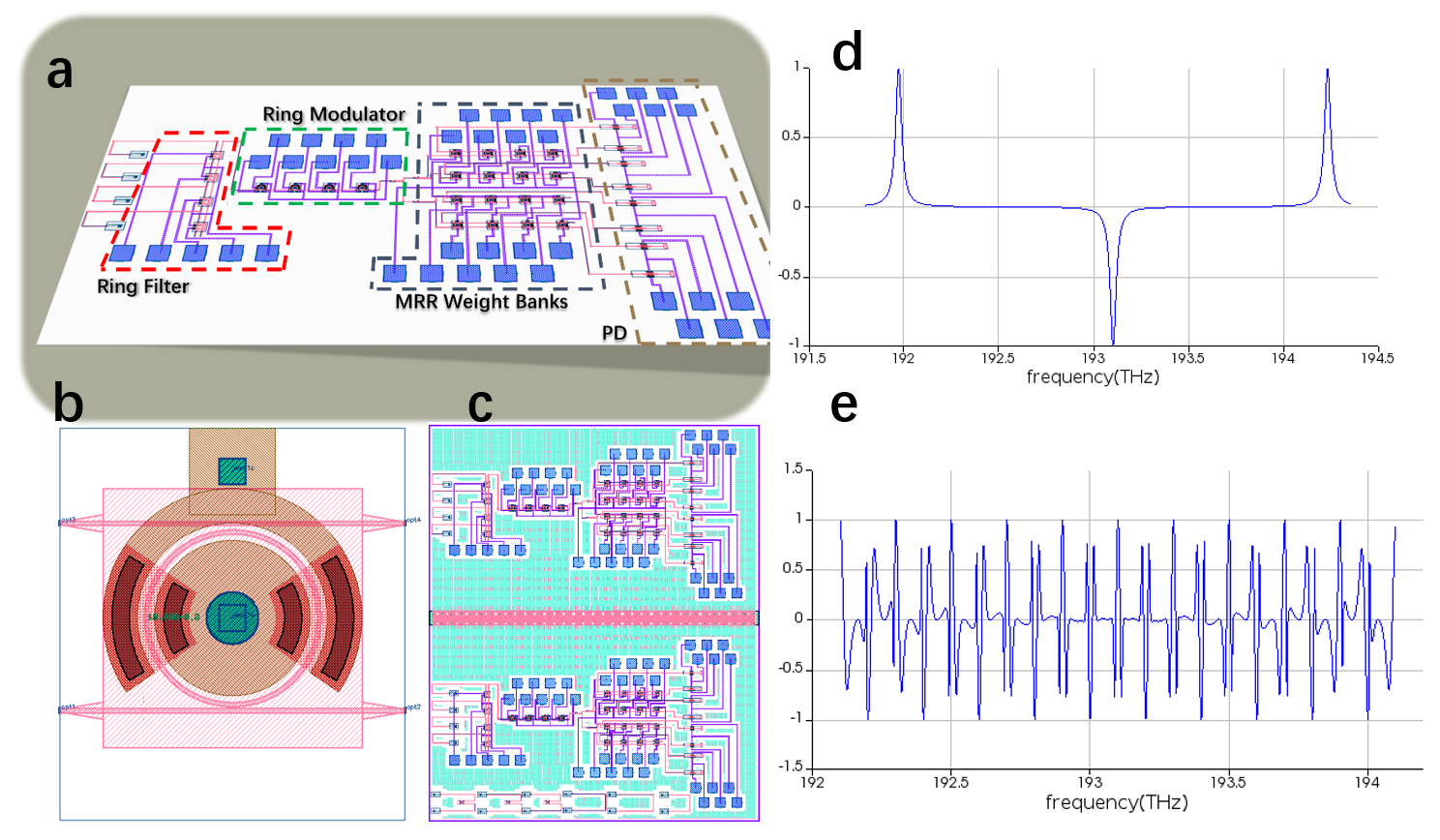}
\caption{\label{fig:circuit} Details in weight bank design. (a). Overview GDS design of the standard weight bank design based on silicon chip. (b). MRR zoomed-in graph with an N-doped in-ring heater. (c) Overall schematic view of weight bank design after tilling. (d). Interconnect outcome of a demux. (e). Interconnect outcome of 4 series ring resonators}	
\end{figure}
The detailed designs of the MRRs are shown in Fig. \ref{fig:circuit}. Each MRR is companied with 2 silicon waveguides with 220$nm$ width which are fully etched. And around MRR is a circular rib waveguide with $90nm$ width which hosts dopants and is slowly etched. To realize feedback control of MRRs, a N-doped photoconductive heaters section near the center of MRR circle is patterned. While a heavier N++ doped section is also patterned for ohmic contacts, according to \cite{ tait2018feedback,jayatilleka2015wavelength}. From \cite{ baehr201225}, the concentration of N \& N++ should be: $ 5* 10^{17} cm^{-3}$, $ 5* 10^{20} cm^{-3}$. The radius of all MRRs and ring filters are all set to 10$\mu m$, but due to the fabrication variance, the final outcome may be different. The coupling gaps between the MRR and rib waveguide is designed to be 200$nm$ on each side. And each ring in neighborhood is isolated by 125$um$.

\begin{table}[ht]

\caption{Definitions of symbols.}
\begin{center}
\begin{tabular}{cc}

\cline{1-2}
\textbf{symbol} &
  \textbf{definition} 
   \\ \cline{1-2}
$\mathrm{\omega_i}$   & Optical carrier frequency    \\ 
$\mathrm{N}$ & Number of optical carriers      \\
$\mathrm{x_i}$ & Data signals          \\
$\mathrm{E_{[in]}(\omega,t)}$  &  Time-frequency expression for WDM input  \\
$\mathrm{E_{0.i}}$ &  Carrier field amplitude   \\
$\mathrm{\delta}$  & Dirac delta function       \\
$\vec{\Delta}$ & Transmission state of filter, resonant wavelength shifts          \\
$\mathrm{H}$ & Tunable spectral filter response   \\
$\mathrm{+/-}$   & Complementary outputs \\
$\mathrm{R(\omega)}$   &  Detector responsivity \\
$\mathrm{\upsilon_{\pi}}$   & Voltage at $\pi$ phase shift \\
$\mathrm{Z_0}$   & Characteristic impedance   \\
$\mathrm{\mu_i}$   & Evaluation of weights          \\
$\mathrm{H^{(+,-)}}$   &  Transmission functions\\        
$\mathrm{T_j}$   &  Amplitude transmission of the MRR through port\\ 
$\mathrm{D_j}$   &  Amplitude transmission of the MRR drop port\\  \cline{1-2}
\label{symbol}  
\end{tabular}
\end{center}
\end{table}

The overview of the standard weight bank design based on silicon chip is displayed in Fig. \ref{fig:circuit}(a). Mental pad arrays around the circuit guides electrical signal from source meter (SM) to individual ring filter and MRR are used to thermally offset resonant frequency of MRR optical transmission to configure its weight \cite{jayatilleka2015wavelength}. The optical wave is directed by grating coupler arrays into and out of ring filter and further flow into MRR weight bank. Both on-chip and off-chip balanced photo-detector (BPD) can be designed to collect the outputs of MRRs by complementary ports and perform electrical weighted addition. MRR zoomed-in graph with an N-doped in-ring heater is also displayed in Fig. 3(b). Due to the inevitable fabrication error that happens in AMF process, we copied the circuit with exactly the same parameter, as is shown in Fig. 3(c), and add several testing circuits to measure the instern loss of the grating coupler for later data analysis. The Interconnect outcome of demux and cascaded four ring resonators are also in fig. 3(d) and (e). 

Physically, this neuromorphic network is implemented by miniature circular waveguide. The corresponding features of neuron nodes are embedded in silicon substrate by nano-scale etching. When input optical signal is captured, MRR weigh bank modulates the output signal of laser that are near threshold. Insignificant disturbance to values which are close to threshold will greatly impact the output signals. WDM is here realized by MRRs, using a specific wavelength of light at nodes in the system. And the non-linearity of the system is realized by feedbacks. MRR weight bank is calibrated with respect to feedback procedure control \cite{ tait2018feedback }. Each MRR is driven by electrical power from probing SM and its resonance can be thermally tuned by corresponding paired laser channel. 

The resonance shift decides circulating optical power within each MRR. The optical power will be in part absorbed which results in a photonic response. The conductivity of the ring-shape N-doped photoconductive heater, on the contrary, will be partially influenced by the photonic response. The conductivity is easy to detect by the probing SM meanwhile \cite{ jayatilleka2015wavelength,orcutt2012open,arakawa2013silicon}. Consequently, by converting sensing the photon response to electronic response, the optical transmission of each MRR can be detected directly, so as to realize the feedback control loop of configuring the weight of MRR in a continuous range.

Assuming that signal bandwidth is much narrower than optical carrier frequency, the time-frequency expression for a WDM input can be concluded by a slowly-varying envelope approximation and short-time Fourier transform \cite{tait2016microring}:

\begin{equation}
E_{[in]}(\omega,t) =
   \sum_{i=1}^{N} E_{0,i} \sqrt{1+x_i(t)} \delta(\omega-\omega_i)
\end{equation}
Where $x_i(t)$ is strictly greater than –1.
The transmission state of filter is configured by tuning a parameter vector, $\vec{\Delta}$.

\begin{equation}
E_{[wei]}^{+,-}(\omega,t) =
   H^{+,-}(\omega; \vec{\Delta}) E_{[in]}(\omega,t)
\end{equation}
+/– superscripts represent complementary outputs of tunable spectral filter response. The effect of a balanced photodiode (PD) is indicated as the difference between two photocurrents derived from the weighted signals.
\begin{equation}
i_{PD} (t) =
   \int_{\omega}R(\omega)(|E_{wei}^+(\omega,y)|^2-|E_{wei}^-(\omega,y)|^2)\, {\rm d}\omega
\end{equation}
Net function fitting the form of weighted addition can be expressed as:

\begin{equation}
\mu_i=
   A_i(|H^+(\omega_i;\overrightarrow \Delta)|^2 - |H^-(\omega_i;\overrightarrow \Delta)|^2)
\end{equation}

\begin{equation}
   y(t) =
   \sum_{i=1}^{N}\mu_ix_i(t)+ sum_{i=1}^{N}\mu_i
\end{equation}
where 
\begin{equation}
A_i\equiv R(\omega_i) \cdot Z_0/\upsilon_{\pi}\cdot E^2_{0,i}
\end{equation}

\begin{equation}
y=i_{PD}\cdot Z_0/\upsilon_{\pi}\cdot E^2_{0,i}
\end{equation}

\begin{equation}
y=\vec{\mu}\cdot\vec{x}
\end{equation}

The final scattering matrix of 2-channel weight bank can be derived:

$$\begin{bmatrix}
\frac{1}{T_1} & 0 & 0 & -D_2T^{-1}_1 \\
0 & \frac{T_1T_2-D_1D_2}{T_2} & D_1T_2^{-1} & 0 \\
0  & -D_2T_2^-1 & T_2^-1 & 0 \\
\frac{D_1}{T_1} &  0 & 0 & \frac{T_1T_2-D_1D_2}{T_1}\\
\end{bmatrix}$$

In order to better understand various symbols and definitions, we summarize all the expressions used in formulas and equations in Table \ref{symbol}.

\section{Conclusion}
\begin{figure}
\centering
\includegraphics[width=0.9\textwidth]{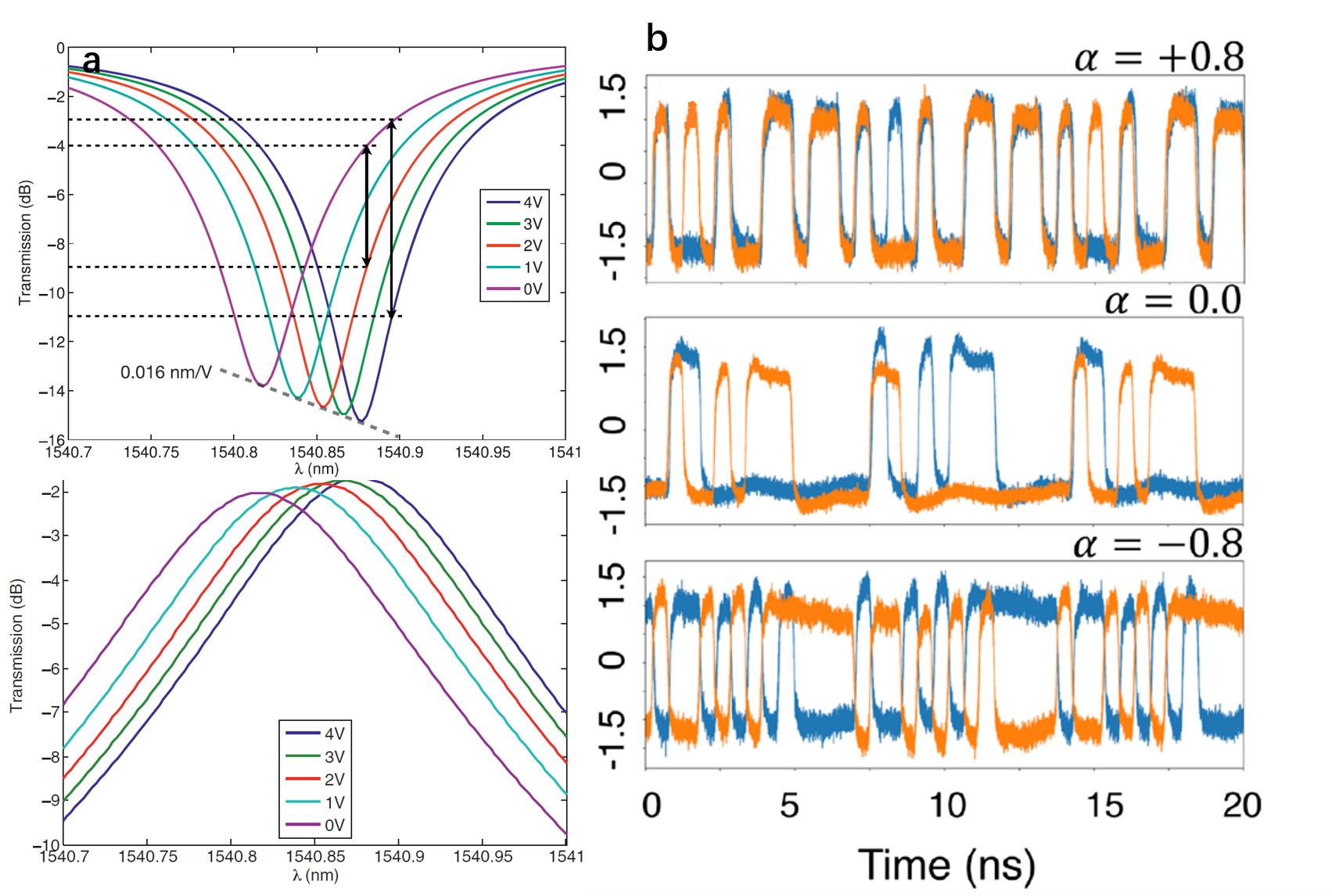}
\caption{\label{fig:expected results} (a). Through port (upper one) and drop port (lower one) spectra with different voltage applied \cite{chrostowski_hochberg_2015}. (b). Waveform pairs of partially correlated signals associated with 3 typical $\alpha $ values ($\alpha =+0.8/1/-0.8$)\cite{chrostowski_hochberg_2015}.}	
\end{figure}

In this project, we designed an on-chip MRR weight bank which implements the photonic addition function. Grating coupler for TE mode wave at 1550nm frequency is used to introduce optical signal into chip. Electrical signal is directed by mental pads to thermally tune each ring filter, ring modulator, MRRs and BPD. In case of fabrication error during chemical process in AMF, we copied the circuit with exactly same parameters to compare the fabrication fitness. And ring-like heaters at ring filters could also help with adjusting the input wavelength center frequency. The mico-rings in the design are implemented with add-drop ring based on reverse-biased PN-junction, where the individual input signal has different center frequency $\lambda_i$ and can only be partially transmitted. And high-Q factor manner of micro-ring will lead to insignificant center spectral shift and massive transmission loss \cite{chrostowski_hochberg_2015} as is show in Fig. \ref{fig:expected results}a. The center spectral shift change due to voltage difference applied to PN-junction indicates transmission function of input signal is dependent on voltage change, which means micro-ring can separately control WDM input pulse transmission through voltage change. Ring modulator together with y-branches of different optical path lengths construct partially correlated inputs for IN port of MRR weight banks. And on-chip BPD is also adapted to perform weight addition/minus operation. Testing circuits with only grating couplers and waveguides are utilized to figure out the insertion loss of GC in future data analysis. After measurement, we expect a result of partially correlated channel signals as is shown in Fig. \ref{fig:expected results}b, which exhibits waveform pairs of partially correlated signals associated with 3 typical $\alpha $ values ($\alpha =+0.8/1/-0.8$) \cite{ ma2019photonic }.

\section{Acknowledgments}
Fabrication support was provided via the Natural Sciences and Engineering Research Council of Canada (NSERC) Silicon Electronic-Photonic Integrated Circuits (SiEPIC) Program and the Canadian Microelectronics Corporation (CMC). Devices were fabricated at Advanced Micro Foundry (AMF) A STAR foundry in Singapore.

\bibliography{sample}

\end{document}